# Universal geometric constraints during epithelial jamming

Lior Atia[1+], Dapeng Bi[2+], Yasha Sharma[1+], Jennifer A. Mitchel[1], Bomi Gweon[2,3], Stephan Koehler[1], Stephen J. DeCamp[1], Bo Lan[1], Rebecca Hirsch[1], Adrian F. Pegoraro[4], Kyu Ha Lee[5], Jacqueline Starr[5], David A. Weitz[4], Adam C. Martin[6], Jin-Ah Park[1], James P. Butler[1]*, Jeffrey J. Fredberg[1]*

[1]Harvard T.H. Chan School of Public Health, Boston, Massachusetts 02115, USA. [2]Northeastern University, Department of Physics, Boston, Massachusetts 02115, USA. [3]Hanyang University, Department of Biomedical Engineering, Seoul 04763, Korea. [4]Harvard University, School of Engineering and Applied Sciences, Cambridge, Massachusetts 02138, USA. [5]The Forsyth Institute, Cambridge, Massachusetts 02142 USA. [6]Massachusetts Institute of Technology, Department of Biology, Cambridge, Massachusetts 02142, USA.

[+]contributed equally; *senior authors

Abstract: **As an injury heals, an embryo develops, or a carcinoma spreads, epithelial cells systematically change their shape. In each of these processes cell shape is studied extensively, whereas variation of shape from cell-to-cell is dismissed most often as biological noise. But where do cell shape and variation of cell shape come from? Here we report that cell shape and shape variation are mutually constrained through a relationship that is purely geometrical. That relationship is shown to govern maturation of the pseudostratified bronchial epithelial layer cultured from both non-asthmatic and asthmatic donors as well as formation of the ventral furrow in the epithelial monolayer of the *Drosophila* embryo *in vivo.* Across these and other vastly different epithelial systems, cell shape variation collapses to a family of distributions that is common to all and potentially universal. That distribution, in turn, is accounted for quantitatively by a mechanistic theory of cell-cell interaction showing that cell shape becomes progressively less elongated and less variable as the layer becomes progressively more jammed. These findings thus uncover a connection between jamming and geometry that is generic –spanning jammed living and inert systems alike– and demonstrate that proximity of the cell layer to the jammed state is the principal determinant of the most primitive features of epithelial cell shape and shape variation.**

Grain in a silo, sand in a pile, and beans in a chute can flow in some circumstances or instead become jammed in others.[1-3] Each of these systems is granular, close-packed, and collective. Moreover, each can exhibit a transition from a fluid-like unjammed phase toward a solid-like jammed phase. The jamming transition has been linked to caging dynamics on the scale of particle-particle interactions, and their associated packing geometries.[4-9]

Cells in a confluent epithelial layer, similarly, can migrate in some circumstances or exhibit migratory arrest in others, with embryonic development, cancer invasion, and wound healing being the classical examples. But compared to that of a close-packed granular collective, the packing geometry and resulting cell shapes that define a confluent epithelial layer would seem to be a different matter altogether.[10-14] We show here, nevertheless, that the epithelial layer defines packing geometries much as do jammed inert collectives. Across vastly diverse epithelial systems, cell shape and the variation of cell shape are shown to be governed by a relationship that is simple and unifying.

To investigate variation of shape from cell-to-cell within the epithelial layer we cultured primary human bronchial epithelial cells (HBECs) obtained from donors who were either non-asthmatic or asthmatic (Methods).[15] Cells were grown to confluence in submerged conditions on a porous transwell for 5-6 days and then allowed to mature in air-liquid interface (ALI) culture conditions to become well-differentiated over the course of 6 to 20 days. Both in ALI culture and *in vivo*, the mature HBEC layer comprises mainly basal cells, goblet cells, and ciliated cells arrayed in a complex pseudostratified structure (Supplementary Fig. 1). In the maturing HBEC layer cell shape is linked to collective migratory dynamics, rates of cellular rearrangement, and cell jamming[15]; the non-asthmatic human HBEC layer jams by day 8 of ALI culture, whereas the asthmatic layer jams no sooner than day 14.[15] Fluorescent images labeled for F-actin showed apical rings of cells tiling the layer (Supplementary Fig. 2-4). These rings were then segmented using a customized algorithm, and from these segmented images we measured cell geometry across all donors and days of maturation in a total of $1.4 \times 10^5$ cells. For each cell we determined the projected area and the aspect ratio, *AR* (Supplementary Fig. 3); the more slender or elongated the cell profile, the higher is its *AR*. For any closed ring in the cell plane, *AR* thus serves as a simple, primitive, and robust metric of shape.

Cell numbers progressively increased and areas progressively decreased with layer maturation (Fig. 1a; Extended Data Fig. 1a,d). In cells from both non-asthmatic and asthmatic donors, cell areas and *ARs* were highly variable (Fig. 1b). Consistent with previous reports[15], cells from asthmatic donors were more elongated than their non-asthmatic counterparts, and the distributions of *ARs* were broad and skewed (Fig. 1c; Extended Data Fig. 1). As a simple measure of shape variation from cell-to-cell, we used the standard deviation of the aspect ratio, SD(*AR*). We had expected this shape variation to represent random biological noise. However, data from all non-asthmatic donors and all days of maturation traced an unanticipated but clear linear relationship between mean of the aspect ratio, $\overline{AR}$, and SD(*AR*) (p<0.0005; Supplementary Table 1); as the cell aspect ratio became progressively smaller with increasing days of maturation, its variation from-cell-to cell did so as well (Fig. 1e). Cells from asthmatic donors mature more slowly and migrate more quickly than do cells from non-asthmatic donors. Therefore we had expected that any relationship between cell shape and its variation might be different as well in non-asthmatic versus asthmatic donors.[15] To our surprise, $\overline{AR}$ and SD(*AR*) from all asthmatic donors and all days of maturation defined a closely similar linear relationship, although with a slightly smaller slope (p<0.0005; Fig 1d; Supplementary Table 1). Moreover, when we pooled data for each day of maturation, $\overline{AR}$ and *SD(AR)* for both asthmatic and non-asthmatics donors traced virtually the same relationship (Fig. 1d). Within this relationship, however, mature and/or non-asthmatic cells tended to fall at lower values of $\overline{AR}$ and SD(AR) whereas immature and/or asthmatic cells tended to fall at larger values (Fig. 1e, Extended Data Fig. 1d).

We considered that the remarkable consistency of this relationship might reflect, 1) a peculiarity of HBECs, 2) an idiosyncrasy of layer maturation, or 3) simply an artifact of cell culture. To address the first possibility, we examined another cell type, the Madin-Darby canine kidney (MDCK) cell. Using Voronoi tessellation based upon nuclear centers, we created a complete polygonal tiling of the cell layer from which we quantified cell AR.[16] In the case of maturing MDCK cells, much as in the case of maturing HBECs, with progressive cell jamming cellular speeds progressively decreased and shape analysis traced the same linear relationship of decreasing $\overline{AR}$ and SD(AR) (Supplementary Video 1), thereby ruling out the first possibility.

To rule out idiosyncrasies of layer maturation, we studied changes in cell shape and shape variation in response to an acute mechanical perturbation; HBECs are known to be mechanosensitive in a manner that contributes to pathological remodeling of the asthmatic airway.[17,18] Therefore, in the mature HBEC layer we applied an apical-to-basal pressure difference (30cmH$_2$O) across the porous transwell. This pressure difference mimics the mechanical compression that occurs *in vivo* during acute asthmatic bronchospasm.[18,19] Such compression squeezes the lateral intercellular space between adjacent epithelial cells, and, through pathways that have yet to be established, ultimately leads to elongated cell shape, increasing cell migration, and unjamming of the layer.[15,17,18] Layer compression caused changes of cell shape and shape variation that tracked from smaller $\overline{AR}$ and SD(AR) to larger (Fig. 1f). Nevertheless, changes in $\overline{AR}$ and SD(AR) caused by layer compression traced the same geometric relationship as did changes observed during layer maturation (cf. Fig. 1e and 1d ). To complement this study in mechanotransduction, we applied sequential stretches to the mature MDCK layer plated on a deformable substrate (6% strain amplitude, 1s duration, once every 6s for 20 min). Immediately after stretch cessation, cell shapes became elongated, but within 60 min these changes relaxed back to pre-stretch values (Fig. 1f). Accordingly, in the cases of both HBEC compression and MDCK stretch, the $\overline{AR}$ and SD(AR) increased in concert to trace the same geometric relationship as observed during layer maturation, but in the opposite sense with time (Fig. 1d,e,f). Moreover, these acute changes could not be accounted for by changes in cell crowding, which were small (data not shown). As such, the geometric relationship traced by the data of Fig. 1d,e,f is not limited to HBECs, and can be attributed neither to epithelial layer maturation nor to changes in cell crowding.

Finally, to rule out artifacts of cell culture and, more importantly, to test the relationship between shape and shape variation in a completely distinct epithelial system *in vivo*, we studied ventral furrow formation in the developing embryo of the fruit fly *Drosophila melanogaster*.[10] In *Drosophila*, the ventral furrow forms to bring mesodermal precursor cells into the interior of the embryo. In this process cells exhibit pulsatile contractions that transition from a relaxed to a sustained contractile state.[20,21] These events are accompanied by changes of cell shape that include apical flattening, constriction of apical diameter, cell elongation, and subsequent shortening [22] (Fig. 2a,b, left column). Across multiple wild type (WT) embryos, we characterized the aspect ratio of cell apical domains and its variability as cells constricted, increased their cellular speed and escaped their dynamically arrested state (Fig. 2c left column), but before they invaginate to form the ventral furrow (Methods). Even though the *Drosophila* epithelium *in vivo* differs in many regards from that of the HBEC and MDCK layers *in vitro,* cell shape data obtained in this system as the furrow forms traced the same geometric relationship (Fig. 2d, left; Supplementary Video 2). To test how shape and shape variation are affected by genetic variation, we studied embryos that were mutant for *concertina* (*cta*). The *cta* gene encodes a $G\alpha_{12/13}$ protein, which is required for coordinated apical constriction.[23] In *cta* maternal effect mutants, cells of the ventral furrow constrict in an uncoordinated manner and furrow formation becomes slow and uneven.[24] In addition, we used RNAi to knock down the transcription factor *twist,* which is essential for sustained apical contraction and ventral furrow formation. Unlike the maturing HBEC layer, but similar to compression-induced unjamming of the HBEC layer, these events in the embryo were accompanied by progressive increases of cellular speeds and $\overline{AR}$s in tandem (Fig. 2c, center and right), and tracked with time from smaller toward larger $\overline{AR}$ and SD(*AR*) (Fig. 2d center and right; Supplementary Videos 3,4).

We found it to be mysterious that cell shapes and shape variation across diverse epithelial systems conform to such a simple and unifying description. Indeed, we knew of no *a priori* reason to expect the existence of a relationship between cell shape variation and cell shape, or that such a relationship would be linear, or that across widely diverse epithelial systems the relationship would be virtually invariant. That these findings cannot be general geometrical properties of any tiled surface is supported by numerous counterexamples, the simplest being any periodic space-filling tiling of the plane, none of which fall close to the relationship. Voronoi tiling of random Poisson seeds generates one datum that does fall on that relationship, interestingly, but that datum falls near the higher, unjammed, end of the biological range (Extended Data Fig. 2). As such, neither totally regular nor totally random tiling of the plane can define the observed geometric relationship or explain it.

To better understand shape variation in each of the cases above, we constructed the corresponding probability density functions (PDFs) of observed *AR*s. During maturation of HBECs or MDCKs, cell shapes became systematically less elongated and less variable with time (Fig. 3a, c). In contrast, during approach to the formation of the ventral furrow in wild type *Drosophila*, cell shapes became more elongated and more variable with time (Fig. 3e) In the *cta* mutant compared with WT, cell shape was more elongated and more variable, and even more so in the *twist* knockdown (Fig. 3e). Despite differences in PDFs within and between these systems, and regardless of their trajectories over time, the functional form of all respective PDFs were unimodal and skewed in a manner suggestive of a common underlying mathematical basis (Fig. 3a,c,e). To assess that commonality in a quantitative fashion, we rescaled *AR* to the form $x = (AR\text{-}1)/(\overline{AR}\text{-}1)$; importantly, $\overline{AR}$ in HBECs progressively fell with increasing days of maturation (p< 0.0005; Extended Data Fig. 1,3), and was systematically greater in non-asthmatic versus asthmatic cells (p<0.0005). Rescaling in this fashion maps the lower limits of the abscissa and the ordinate to zero, and scales the mean of the distribution to unity. Across all days of maturation for HBEC data in all non-asthmatic donors, data thereupon collapsed onto a common distribution (Fig. 3a,b). For asthmatic donors, too, rescaled data again collapsed to a common distribution (Fig. 3a,b), as did data from MDCK cells (Fig. 3c,d), events preceding formation of the ventral furrow, and interventions that disrupt that furrow formation (Fig. 3e,f). Within and across these vastly diverse epithelial layers, observed variation of cell shape was well-described by a single distribution that, with small distinctions, was common to all and, potentially, universal.

We wondered, therefore, if these findings might be explained by a generic connection between jamming and geometry that is not restricted to the particular effects of cell crowding, cell adhesion, or even system dimensionality, and instead points to a larger jamming reference class that transcends such system details (Fig. 4; Extended Data Fig. 2). To explore that possibility we start by considering the densely-packed, 3D, jammed, inert, granular collective, wherein grain centers can be used to tessellate the volume around each individual grain. When packing is dense and random, these tessellated volumes necessarily vary from grain to grain. In such a system, local variations of available volume, *x*, have been found to follow the *k*-gamma distribution[25],

$$\text{PDF}(x; k) = k^k x^{k-1} e^{-kx}/\Gamma(k) \qquad (1)$$

where $\Gamma(k)$ is the Legendre gamma function. This distribution is fully described by only a single parameter, *k*, and has a mean of unity. A rigorous basis for applicability of Eq. 1 to the confluent cell collective remains to be established, of course. Nevertheless, if we take $x = (AR\text{-}1)/(\overline{AR}\text{-}1)$ and use maximum likelihood estimation (MLE; Methods) to determine numerical values of *k*, Eq. 1 is then seen to be faithful to observations with a high degree of statistical confidence (Extended Data Fig. 3,4). In HBECs, the parameter, *k*, varied little over all days of maturation (p<0.0005, Extended Data Fig. 3) and, moreover, did not differ between non-asthmatic versus asthmatic cells (p=0.2147, Extended Data Fig. 3). Even though some differences in *k* were statistically significant, these differences were at most modest and, to a reasonable approximation, *k* in HBECs could be taken as a constant value of 1.97 (95% CI [1.89, 2.04]). In MDCK cells we found, similarly, that all data were well represented by a single *k* value of 2.31 (CI [1.90, 2.73]). In *Drosophila, k* values were not statistically different between WT and *cta* mutant (p=0.023), or *twist RNAi* (p =0.762) (Extended Data Fig. 4), and were well represented by a single *k* value of 2.52 (CI [2.43, 2.62]). Across these diverse epithelial systems, Eq. 1 pertained throughout and *k* was bounded to the relatively narrow range between 2 and 2.5.

For granular matter, the mathematical form of the *k*-gamma distribution (Eq. 1) links volume variation to mean volume according to the equation $SD(V) = (\overline{V} - V_{min})/\sqrt{k}$, where *V* represents local tessellated volume associated with each individual grain, $\overline{V}$ is the mean value of such volumes, and $V_{min}$ is the minimal random packing volume that the system can attain.[25] But just as Eq. 1 requires volume variation $SD(V)$ to change linearly with mean volume $\overline{V}$ in the case of granular systems, so too Eq. 1 requires cell shape variation $SD(AR)$ to change linearly with mean cell shape $\overline{AR}$ in the case of epithelial systems (Fig. 1d, 2d). Furthermore, Eq. 1 derives from maximization of distributional entropy, $S$.[25] Because volume in an athermal granular system plays a role analogous to that of energy in a thermodynamic system, one can then define an effective temperature, $T_{eff}$, given by $1/(\partial S/\partial V)$.[25] When carried over from volume variation in granular jamming to shape variation in cellular jamming, $T_{eff}$ then takes the form,

$$T_{eff} = SD(AR)/\sqrt{k} = (\overline{AR} - AR_{min})/k. \qquad (2)$$

To the extent that Eq. 1 pertains (Fig. 3b,d,f) and to the extent that observed changes in *k* across epithelial systems are small (Extended Data Fig. 3b, 4b), Eq. 2 would then explain not only the existence of a relationship between cell shape variation and cell shape, but also its linearity and its invariance.

From the jamming analogy a further insight follows. The numerical value of *k* in 3D granular jamming has been shown to represent the number of elementary tessellated regions, or an effective cluster size, with which a typical region can mutually exchange space.[25] Across a diverse range of such inert granular systems, the experimentally determined value of *k* is found to be robust and close to 12, thus suggesting that each grain in 3D interacts with roughly 12 nearest neighbors. To the extent that such an interpretation pertains to shape variation within the jammed epithelial layer, data described above implies that cell shape within the epithelial layer interacts with only 2 to 3 nearest neighbors, and, as such, that cell-cell shape interactions are predominantly short-ranged.

This physical picture finds direct support in computational simulations of cell-cell mechanical interactions. The cell jamming model of Bi *et al.* proposes that there exists a preferred cell shape that is set by a competition between cell cortical contraction and cell-cell adhesion (Extended Data Fig. 5).[12,26-29] Contraction versus adhesion are in competition because the former acts to reduce the length of each cell-cell junction whereas the latter acts to increase it. When cortical tension dominates, the collective is predicted to become solid-like and jammed. As a result, each cell becomes trapped in a shape that departs from its preferred shapes, and $\overline{AR}$ becomes frozen at a value close to 1.2 (Extended Data Fig. 5). As cell-cell adhesive effects progressively increase (or, equivalently, as cortical contractile effects progressively decrease), cell shapes remain trapped until some critical condition is ultimately reached. At that critical point the collective undergoes a transition from a solid-like jammed phase to a fluid-like unjammed phase. Cells within the layer thereupon become increasingly free to attain their preferred shapes, with both $\overline{AR}$ and the variation in AR increasing as the cell layer becomes progressively more and more unjammed (Extended Data Fig. 5). For a wide range of computational cases, this approach provides independent confirmation of: 1) a linear relationship between SD(AR) and $\overline{AR}$, much as in Eq. 2 and Fig 3g (inset), 2) a range of probability density functions all of which are concordant with the k-gamma distribution, much as in Eq. 1 and Fig. 3g, and, as determined from MLE analysis, 3) collapse of those predicted probability density functions to a universal distribution characterized by a single numerical value of k (Fig. 3h). And although this computational approach is in many regards overly simplistic, it makes a quantitative prediction concerning the numerical value of k (2.53), a value that slightly overestimates observations in HBECs and MDCKs but is in accord with observations in *Drosophila*.

Experimental observations of cell shape spanning vastly diverse epithelia, the statistical physics of disordered granular matter, and computational mechanics, taken together, are thus seen to converge in manner that is qualitatively consistent and quantitatively reinforcing. This unification across living versus inert jammed matter (Fig. 4) provides an insight that is to us as novel as it is striking.

Cell shape has long intrigued epithelial biologists. D'Arcy Thomson noted the similarity between the structure of the epithelial cell collective and a soap foam.[30] That observation eventually led to the honeycomb conjecture, which holds that a regular hexagonal tiling corresponds to optimal cell packing at minimum material cost.[31] Cell divisions and apoptosis are known to drive cell arrangements away from such hexagonal packing and toward an equilibrium distribution of cellular polygons.[14] Nevertheless, shape distributions measured throughout the jamming process of HBECs depart systematically from that predicted equilibrium distribution (Extended Data Fig. 1e). Moreover, cell divisions tend to orient along the long axis of the interphase cell in a manner that facilitates both stress relaxation and isotropic growth within the cell layer[32] –a phenomenon commonly known as Hertwig's rule.[33] In setting cell shape and shape variation, the jamming concept does not dispute roles for proliferation, crowding, extrusion, apoptosis, or even newly discovered nematic defects[14], but rather subsumes them into a mechanistic framework that is overriding. For example, cell division or apoptosis[11] are possible mechanisms that can drive changes in cell shape and a corresponding transition between a solid-like and a fluid-like state in a process that we would now call unjamming.[34] Although open questions and unresolved issues remain (Supplementary Information 5), these findings suggest that the most primitive metrics of cell shape –the cell aspect ratio and its variation from cell to cell– are linked to one another and, together, are set by proximity of the layer to the jammed state (Supplementary Videos 1-4).

Data obtained across vastly different epithelial systems emphasize the generality of this concept. But to the extent that cell jamming imposes a physical constraint upon cell shape variation, one might ask, how did cell jamming come into play in the earliest multicellular aggregates? And what biological function might it serve? It has been argued that in order to create multicellular aggregates with the capacities to elongate, fold, segment, or migrate, early evolutionary events must have harnessed, bootstrapped, or otherwise worked around physical processes generic to materials that are condensed, soft, excitable, and viscoelastic.[35] One such physical process is cell jamming and, in support of that notion, the epithelial layers reported here are seen to define microscale geometries much as do jammed close-packed inert collectives (Fig. 4). Through the lens of such a physical perspective, jamming and unjamming might thus be seen as primitive modules that early multicellular organisms could not escape[36], with unjamming activated to enable wound healing, development, or invasion, and jamming being harnessed to slow those processes or arrest them.

In this report physics and biology come together to show that cell shape and shape variation across a wide range of epithelial systems are signatures of the cell jamming process and are fully accounted for by that process, both qualitatively and quantitatively. However, cell jamming in no way challenges the importance of underlying gene regulation, cell effectors, or tissue-scale mechanics, which have been called the troika of tissue shaping during morphogenesis.[10] Rather, cell jamming suggests a specific mechanism at an intermediate scale through which such molecular scale events might be expressed to cause tissue-level outcomes. For example, it has been well-recognized that each component of the troika cannot exert its effects upon cell shape directly, but rather must mediate its effects through the combined but indirect actions of genetic, cellular and mechanical inputs that remain incompletely understood.[10] Here we build directly upon this idea and extend it. To set the most primitive features of cell shape and shape variation, these indirect actions must combine to modulate proximity of the confluent cell layer to the jammed state, and thereby shift the layer up or down the geometric relationship shown in Fig. 1-3. But they do not cause the confluent layer to depart from this relationship. On the scale of cell dimensions, therefore, these findings support the cell jamming process as the principal determinant of epithelial cell shape and shape variation.

Acknowledgements: The authors thank M. Lisa Manning and Emil Millet for helpful discussions. This work was funded by the National Cancer Institute grant number 1U01CA202123 and by the National Heart Lung and Blood Institute grant numbers R01HL107561, PO1HL120839, and T32 HL007118, and the National Research Foundation of Korea grant number NRF-2014R1A6A3A04059713.

## Methods

**Culture of primary HBECs in ALI culture:** Primary human bronchial epithelial cells (HBECs) were obtained at passage 0 or passage 1 from the Marsico Lung Institute/Cystic Fibrosis Center at the University of North Carolina, Chapel Hill. HBECs were cultured from three non-asthmatic and three asthmatic donors as previously described.[15] Briefly, passage 2 HBECs were seeded onto a transwell insert coated with type I collagen (2 transwells per donor), and grown under submerged conditions for five to six days until the cells reached confluence. Upon reaching confluence, media was removed from the apical side of the transwell, but was kept in the basal side to initiate air-liquid interface (ALI) culture conditions. Cells were maintained in ALI conditions and became well-differentiated, expressing basal, goblet and ciliated cells (Supplementary Fig. 1), as seen in airways *in vivo*. On specific days of ALI (Fig. 1) cells were fixed with 4% paraformaldehyde (PFA) and stained with phalloidin conjugated with alexa-488 to visualize F-actin (Life Technologies). Wide field fluorescent images (10-20 per transwell) were acquired at the apical plane on a Leica DMI 8 microscope using either a 40X or 63X oil objective (Leica), and automatically segmented and analyzed using an in-house custom algorithm (Supplementary Methods).

**MDCK culture and live imaging:** Cells were stably transfected with GFP-linked nuclear localization (NLS-GFP) and cultured in Dulbecco's modified Eagle's medium (DMEM) supplemented with 10% fetal bovine serum (FBS), 1% penicillin/streptomycin, and 0.5 mg/ml G418. Tissue culture plates were coated with collagen I and 50,000 cells were added to the center of the well and allowed to adhere for 24 hours. Cells were maintained at 37C and 5% $CO_2$ as images were recorded every three minutes for 54 hours using phase microscopy and confocal fluorescence using the 488 nm line of an argon laser on a Leica DMI6000 SP5 microscope with a stage top incubator. Images were processed to find centroids of each nuclei, which were used as seeds for a Voronoi tessellation in order to create a polygonal tiling of the cell layer. These polygons were then used to obtain cell aspect ratio using an in-house custom algorithm.

**Fly stocks and live imaging:** The following fly lines were used for imaging: *sqh*::*GFP* (myosin regulatory light chain, *sqh*, fused to GFP and expressed from endogenous promoter) (PMID: 12105185), *Gap43-mCherry* (membrane marker driven by *sqh* promoter) (PMID: 20194639). To make maternal effect *concertina* (*cta*) mutant embryos, the $cta^{RC10}$; *sqh::GFP, Gap43::mCherry/TM3* flies were crossed with *Df(2L)PR31/CyO, sqh::GFP* flies and embryos from nonbalancer F1 females were imaged.

For live-cell imaging of *Drosophila*, embryos were dechorionated with 50% commercial bleach, washed with water, and mounted ventral side up on a glue-coated microscope slide. Two No. 1.5 coverslips were put between the slide and top coverslip to make a chamber and avoid compressing the embryo.

Embryos were imaged in halocarbon 27 oil (Sigma-Aldrich). Images were acquired using a Zeiss LSM 710 confocal microscope equipped with a x 40/1.2 numerical aperture Apochromat water objective (Carl Zeiss). For each time point, z-projections of three consecutive slices were automatically segmented and analyzed using an in-house custom algorithm (Supplementary Methods).

**RNAi knock-down in *Drosophila*:** To disrupt the *twist* transcription factor, we injected dsRNA generated using the Invitrogen MEGAscript T7 transcription kit, resuspended in 0.1 x PBS (PMID: 19029882). The dsRNA against *twist* had to be injected at least 2.5 hours before imaging/gastrulation in order to observe the phenotype. The following primers were used to generate *twist* dsRNA: F: 5'-TAATACGACTCACTATAGGGGCCAAGCAAGATCACCAAAT-3'; R: 5'-TAATAC GACTCACTATAGGGGACCTCGTTGCTGGGTATGT-3'.

**Maximum likelihood estimation (MLE):** We fitted the k-gamma probability density, PDF$(x; k) = k^k x^{k-1} e^{-kx}/\Gamma(k)$, to the data by the method of maximum likelihood. For each data set, aspect ratios were shifted and normalized by $x = (AR-1)/(\overline{AR}-1)$. For a given set of data $\{x_i\}\ i = 1,...,N$, the likelihood function is,

$$L(k) = \prod_{i=1}^{N} \rho(x_i; k)$$

The MLE estimate of $k$ is then given by $\hat{k} = \text{argmin}|_k (-\ln L(k))$. The confidence intervals for $\hat{k}$ were determined by parametric bootstrapping as follows. $N$ simulated data points were generated from the cumulative distribution function (CDF) corresponding to the $\hat{k}$ for that set. Maximum likelihood was again performed, yielding a parameter estimate $\hat{k}_1$ from this first set of simulated data. This procedure was repeated over a large number $S$ of simulations, typically >200. The standard deviation of the resulting set $\{\hat{k}_s\}$, $s = 1,...,S$ was used to construct 95% confidence intervals.

Finally, we tested goodness-of-fit for this approach by computing Chi squared values. For each data set, we constructed (uniform) decile bins from the cumulative distribution function, given $\hat{k}$ for that set. We counted the observed number of data points $O_j$, $j = 1,...,10$ occurring within each bin, and compared these with the expected number of occurrences $E = N/10$ within each bin by computing,

$$\chi^2 = \sum_{j=1}^{N}(O_j - E)^2 / E.$$

For decile binning, the number of degrees of freedom is 9; the threshold value of $\chi^2$ for 95% confidence is given by $\chi^2_{0.05,9} = 16.9$. Observed values of $\chi^2 > \chi^2_{0.05,9}$ indicate a rejection of goodness-of-fit with 95% confidence.

For HBEC data, we found that out of 646 frames only 22 had $\chi^2 > \chi^2_{0.05,9}$, or approximately 3.4%. This is close to the expected false negative rate of 5%, given an appropriate distribution function with 95% confidence limits. Similarly, the data from Drosophila revealed a fraction of $\chi^2 > \chi^2_{0.05,9}$ also approximately 3.4%, and the MDCK data exhibited a fraction 9.8%. Taken together, this is striking confirmation at the level of goodness-of-fit that the underlying distribution over these cell types is well characterized by the *k*-gamma distribution given in Eq. 1.

# Figures and legends

## Figure 1

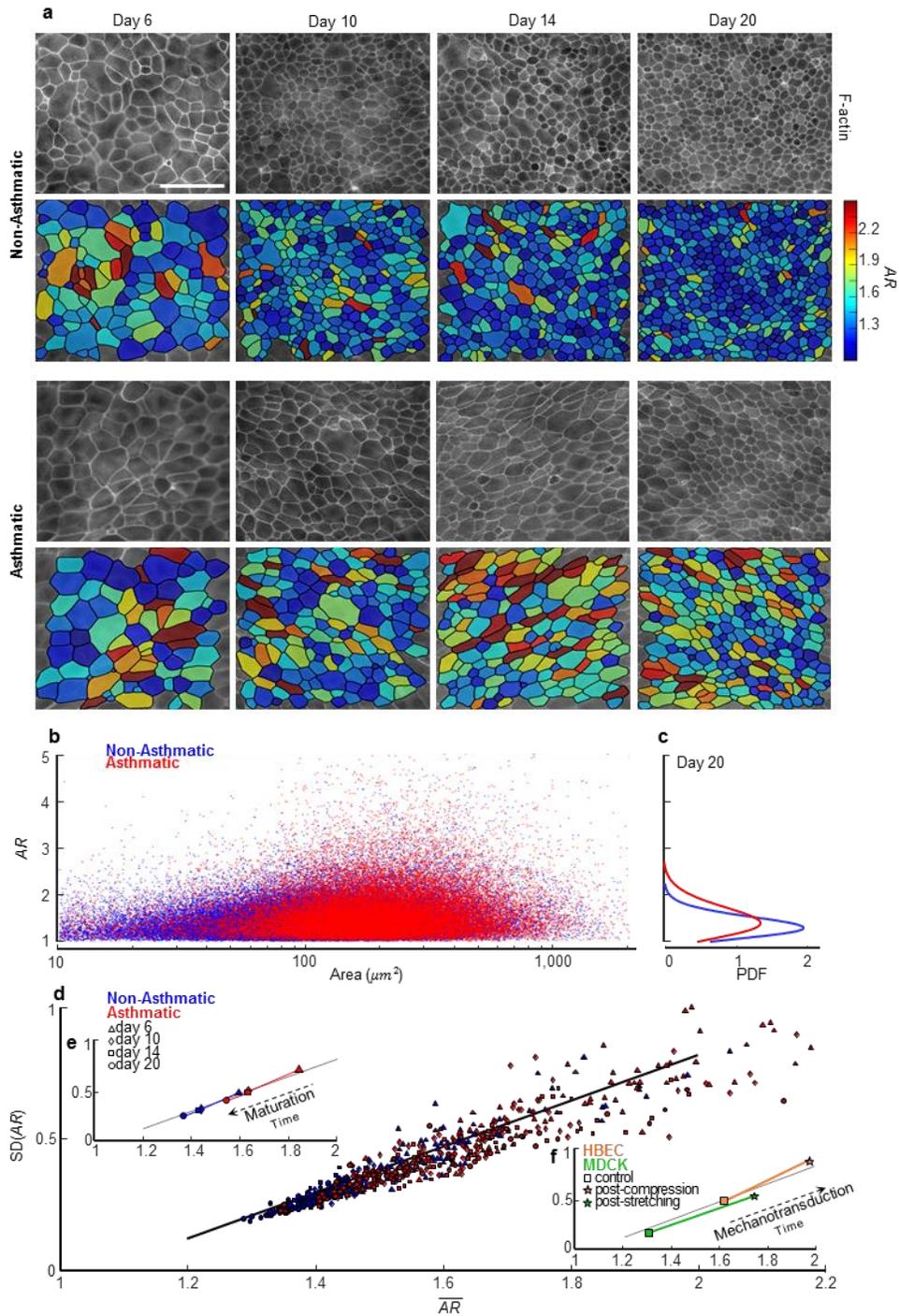

**Figure 1| Across non-asthmatic and asthmatic donors of primary human bronchial epithelial cells (HBECs), and across all days of maturation, cell shape and shape variation *in vitro* are mutually constrained.**
**a,** The apical actin ring was used to measure projected cell area and cell aspect ratio (*AR*; Supplementary Fig. 2-4). With ongoing layer maturation, the *AR* of both non-asthmatic and asthmatic cells became progressively smaller and less variable (corresponding color maps). Scale bar, 50 µm.

**b,** In cells from both non-asthmatic donors (n=87,066 cells, blue) and asthmatic donors (n=46,076 cells, red), *AR* was highly variable but did not co-vary with projected cell area.

**c,** In the mature layer (20 days), the distributions of *AR* in both non-asthmatic and asthmatic cells were wide and skewed.

**d,** *AR* from all donors and all days of maturation defined a clear relationship between the mean of *AR* ($\overline{AR}$) and the standard deviation (SD) of *AR* (each datum represents a different field of view), with non-asthmatic cells (day 6, 10, 14, 20; correspondingly n=13293, 23624, 20371, 29778) tending to fall at lower values of $\overline{AR}$ and SD(*AR*), and asthmatic cells (day 6, 10, 14, 20; correspondingly n=5372, 12629, 10019, 18056) tending to fall at larger values.

**e,** With increasing days of maturation $\overline{AR}$ and SD(*AR*) decreased in tandem (each datum pools all cells for a given day of maturation), but were systematically increased in asthmatic compared with non-asthmatic controls. Nevertheless, all observations fell onto the same relationship.

**f,** Application of an apical-to-basal pressure difference simulates bronchospastic compression of the layer[18] and causes layer unjamming.[15] Corresponding changes in $\overline{AR}$ and SD(*AR*) (Each datum pools all cells for a given sample) tracked along the same geometric relationship (control, n=6153; post-compression, n=5659) as did maturational changes, but in the opposite sense in time. Similarly, the mature MDCK layer subjected to stretch tracked along the same relationship (control, n=367; post-compression, n=422).

The dark continuous line (**d**,**e**,**f**) is not a regression line, but rather the prediction given by the computational model (Extended Data Fig 5). A linear regression for all data in **d** gives SD(*AR*)=0.808 $\overline{AR}$ -0.85 ($R^2$=0.878).

# Figure 2

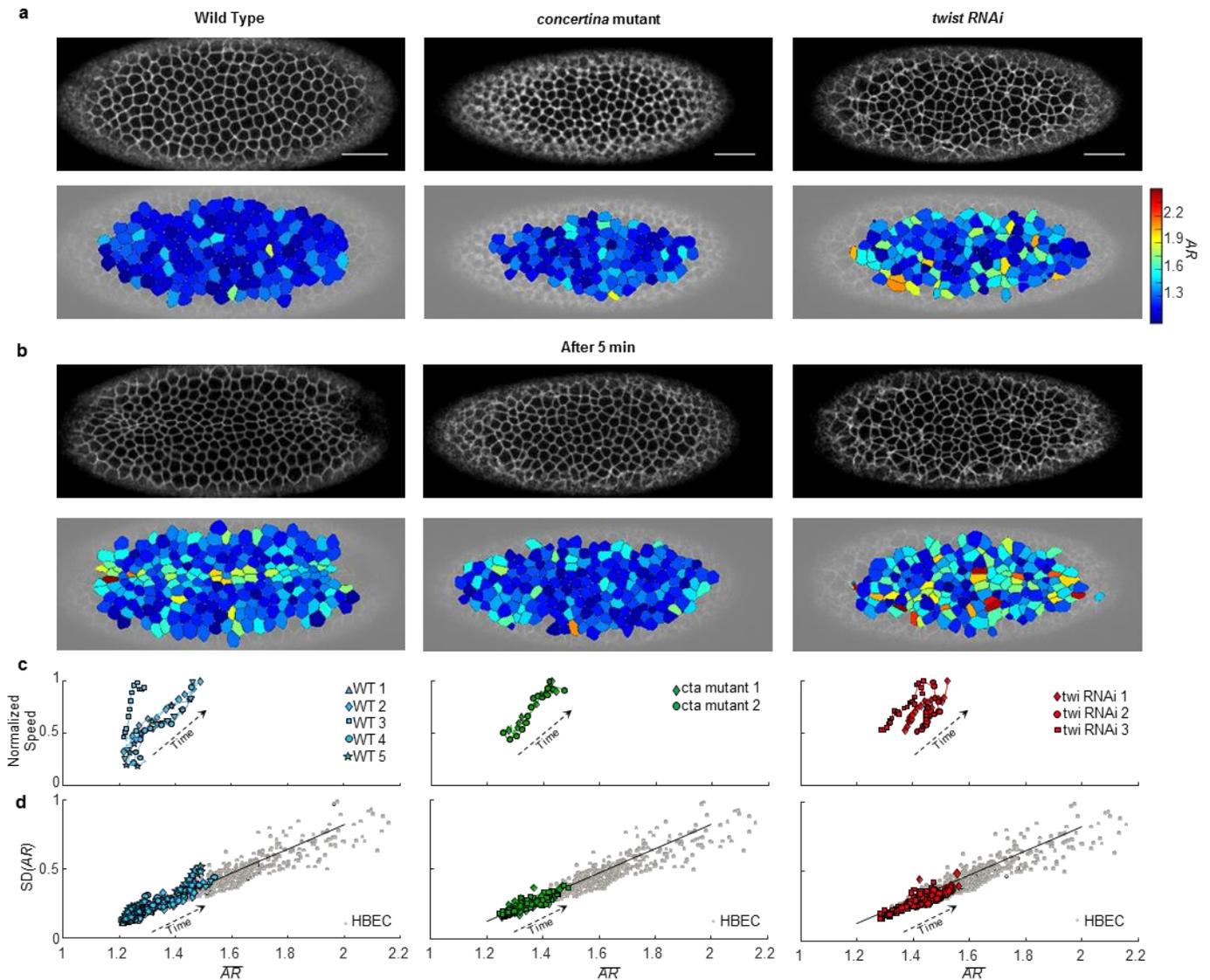

**Figure 2| During ventral furrow formation in *Dropsophila in-vivo,* cell shape and shape variation follow the same geoemtrical relationship as do HBECs *in vitro*.**

**a**,**b,** During *Drosophila* gastrulation, a membrane marker was used to identify cell-cell boundaries with ongoing formation of the ventral furrow in Wild Type (WT), *concertina (cta)* mutant, and *twist* (twi) RNAi embryos (n=5, 2 and 3 respectively). Aspect ratio (*AR*) was measured for 100-150 cells per frame, for 60-100 frames (6-8 seconds apart) per embryo. With time (**b**, after 5 minutes) cells constricted and attempted to form the ventral furrow, and cell *AR* in all embryos became progressively larger and more variable (corresponding color maps). Scale bar, 25 μm.

**c,** Average speed (across all cells) and $\overline{AR}$ increased in tandem for WT, *cta* mutant and twi-RNAi embryos. The speed is normalized to the maximum speed observed in each embryo, typically corresponding with the initiation of the fold, and ranging from 3 μm/min in WT to 1.5 μm/min in *cta* mutant and *twist* (twi) RNAi embryos.

**d,** With time, $\overline{AR}$ and SD(*AR*) increased in tandem in WT embryos (**c**, left) and followed the same geometric relationship as seen in HBECs *in-vitro* (Fig. 1d,e,f; Supplementary Video 2-4).This relationship persisted in embryos with genetic variation (**c**, center and left) that prevent or hinder the ventral furrow formation, as in *cta* mutant and *twist* (twi) RNAi embryos respectively. (Each datum represents a frame from an individual embryo). Each gray datum represents a different field of view for both asthmatic and non-asthmatic HBEC. The dark continuous line is the same as in Fig. 1, and corresponds to the prediction given by the computational model (Extended Data Fig. 5).

**Figure 3**

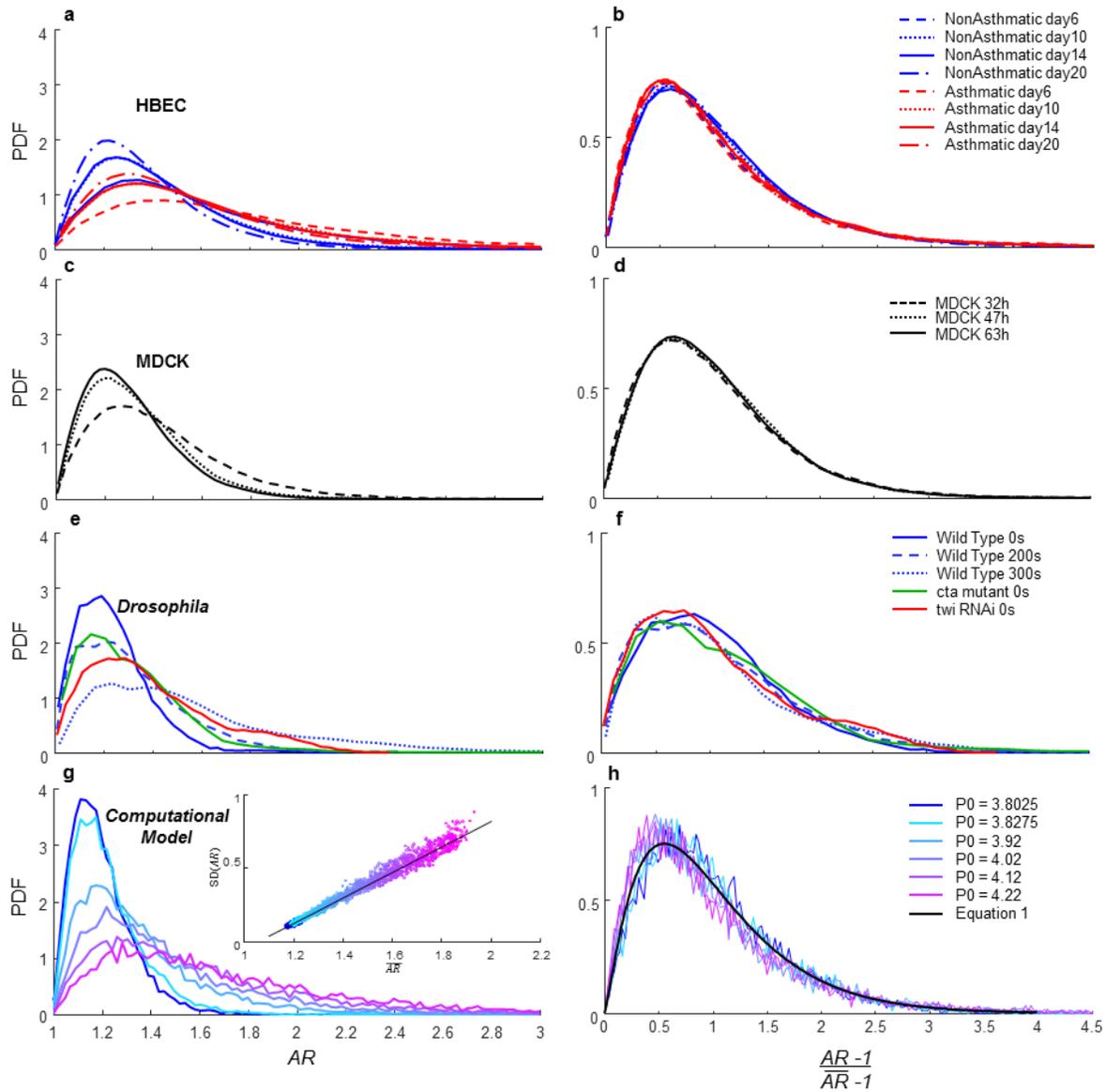

**Figure 3| Within and across vastly different epithelial systems, shape variation collapses to a family of probability density functions (PDFs) that is common to all, and perhaps universal.**

**a,** PDFs of *AR* in HBECs (Fig. 1d,e) became systematically less skewed and less variable with maturation.

**b,** PDFs of the rescaled parameter $x = (AR-1)/(\overline{AR}-1)$ of HBECs, where $\overline{AR}$ denotes the mean *AR* for each respective distribution, followed a *k*-gamma distribution (Eq.1) with *k*= 1.97.

**c,d,** Similar to **a,b** respectively for MDCKs with k = 2.31.

**e,f,** Similar to **a,b** respectively for *Drosophila* (Fig. 2d) with k = 2.52.

**g** Predicted distributions for *AR* given by the model of Bi *et al.*[26,27] Inset shows predicted relationship for $\overline{AR}$ vs. SD(*AR*) (Extended Data Fig. 5).

**h,** Collapse of predicted distributions. Black line shows maximum likelihood estimation (MLE) fit with k = 2.53.

**Figure 4**

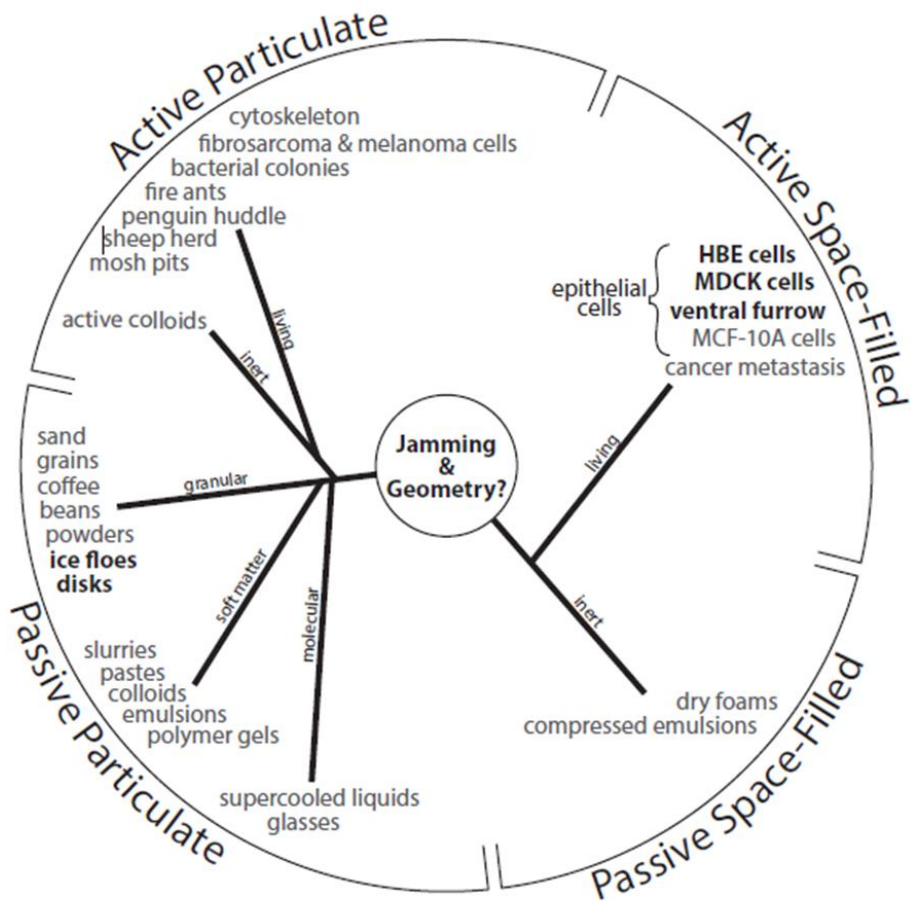

**Figure 4| The jamming superfamily**. Each entry denotes a phyical system in which jamming behavior has been reported, and is categorized here into four classes. In bold are systems described in this report; See also Extended Data Fig. 2. The jamming mechanism was first introduced to explain the poorly understood behaviors that typify certain collective granular systems. Similarities in behavior between such inert granular systems and the migrating epithelial layer were quickly recognized however.[37-39] For example, both granular and cell collective systems are close-packed, volume exclusion prevents two particles (or cells) from occupying the same space at the same time, and particle-particle (cell-cell) interactions are strong. Moreover, just as inert granular systems display swirling motions that arise in cooperative multi-particle packs and clusters, so too does the migrating epithelial layer.[4,40-44] But other physical factors do not fit so easily into this analogy. For example, within granular matter the state of internal mechanical stress is mainly compressive –these are fragile materials in the sense that they can support immense compressive stresses but can support no tensile stress whatsoever– whereas within the confluent cell layer the mechanical stress is overwhelming tensile.[43] Within granular matter a principal control variable for jamming is free space between grains[2,45] whereas in the fully confluent cell layer there is by definition no free space between cells. Within granular matter neither a change of particle shape nor mutual particle-particle adhesion is required for jamming or unjamming – although either can influence jamming dynamics[46]– whereas cell shape change and cell-cell adhesion are thought to be indispensable features of epithelial function and jamming.[26,27,34,47] And perhaps most importantly, the granular particle is neither active nor self-propulsive nor mechanosensitive whereas the epithelial cell exhibits all of these characteristics. In this report we show that the behavior of these diverse living and inert systems is unified to a remarkable extent by consideration of system geometry.

# Extended Data

**Extended Data Figure 1**

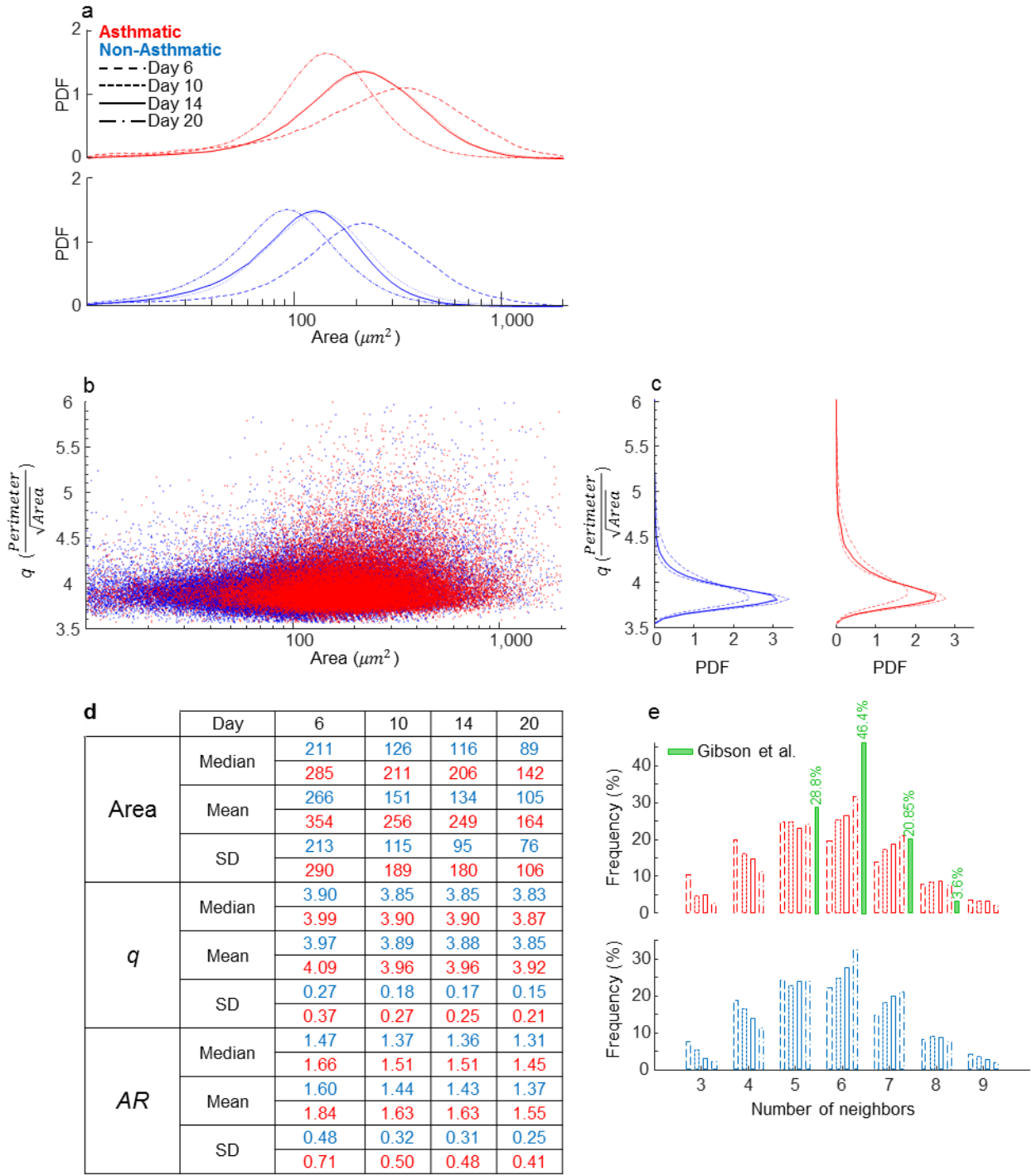

**Extended Data Figure 1| Distributions of cell size and shape metrics differ between asthmatic and non-asthmatic HBECs, and are not accounted for by the equilibrium distribution of polygons predicted for proliferating epithelia**.
**a**, Across non-asthmatic and asthmatic HBECs, and across all days of maturation, cell area was broadly distributed. As cell numbers proliferate with layer maturation these distributions shift to the left (Fig. 1**a**, 1st and 3rd rows). For each ALI day, cell area of asthmatic HBECs was greater than for non-asthmatics.

**b**, In both asthmatic (n=46,076 cells) and non-asthmatic (n=87,066 cells) HBECs, the shape metric q (perimeter/$\sqrt{area}$) was highly variable but did not co-vary with projected cell area (compare Fig.1b).

**c**, PDFs of *q* became systematically less skewed and less variable with maturation. For every given ALI day, *q* of the asthmatic HBECs was higher than for the non-asthmatic ones.

**d**, Median, Mean, and standard deviation (SD) of both cell area and cell shape metrics, *q* and *AR*, became progressively smaller with increasing days of maturation and their respective distributions became less variable (**a**, **c**, Fig. 3a). Consistent with previous report, at day 20 of ALI the non-asthmatic cells attained a median *q* value of 3.83, similar to the critical *q* value associated with jammed epithelial layers.[15]

**e**, Most cells had 5 or 6 immediate neighbors, although the number of nearest neighbor was often as small as 3 or as large as 9. With layer maturation the distribution of nearest neighbors became somewhat more tightly clustered around 6, but did not differ between non-asthmatic and asthmatic HBECs. These distributions were substantially wider than the predicted nearest-neighbor equilibrium distribution in proliferating epithelia from Gibson et al.[11] , shown in green.

**Extended Data Figure 2**

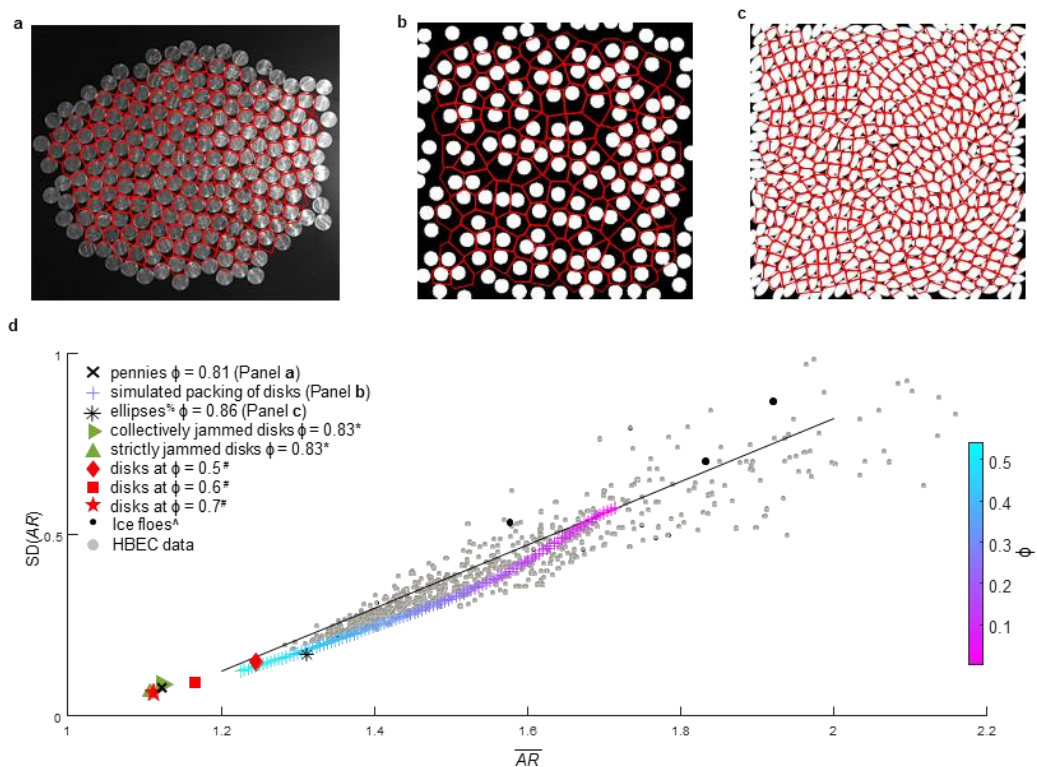

**Extended Data Figure 2| Jamming inert granular 2D systems conform to the same packing geometries as do HBECs.** Using Voronoi tessellation based upon centers of constituent particles in a variety of systems at different packing fractions (ϕ), we created a complete polygonal tiling of the occupied surface to generate equivalent cells. Cell aspect ratio (*AR*) was measured (Supplementary Fig. 3) to obtain a relationship between the variability of cell *AR* and its mean in these non-confluent systems.
**a,** Pennies randomly packed on a surface (ϕ = 0.81).
**b,** Random sequential addition of disks[48]; ϕ approaching zero corresponds to random tiling of Poisson seeds.
**c,** Tightly packed ellipses (ϕ = 0.86).[49]
**d,** *AR* from all systems defined a clear relationship between the standard deviation (SD) of *AR* and its mean, independent of ϕ. Each gray datum represents a different field of view for both asthmatic and non-asthmatic HBEC. In addition, other systems of disks with higher packing fraction *[6,50] and ice floes from the Arctic region follow the same relationship. ^[51]

**Extended Data Figure 3**

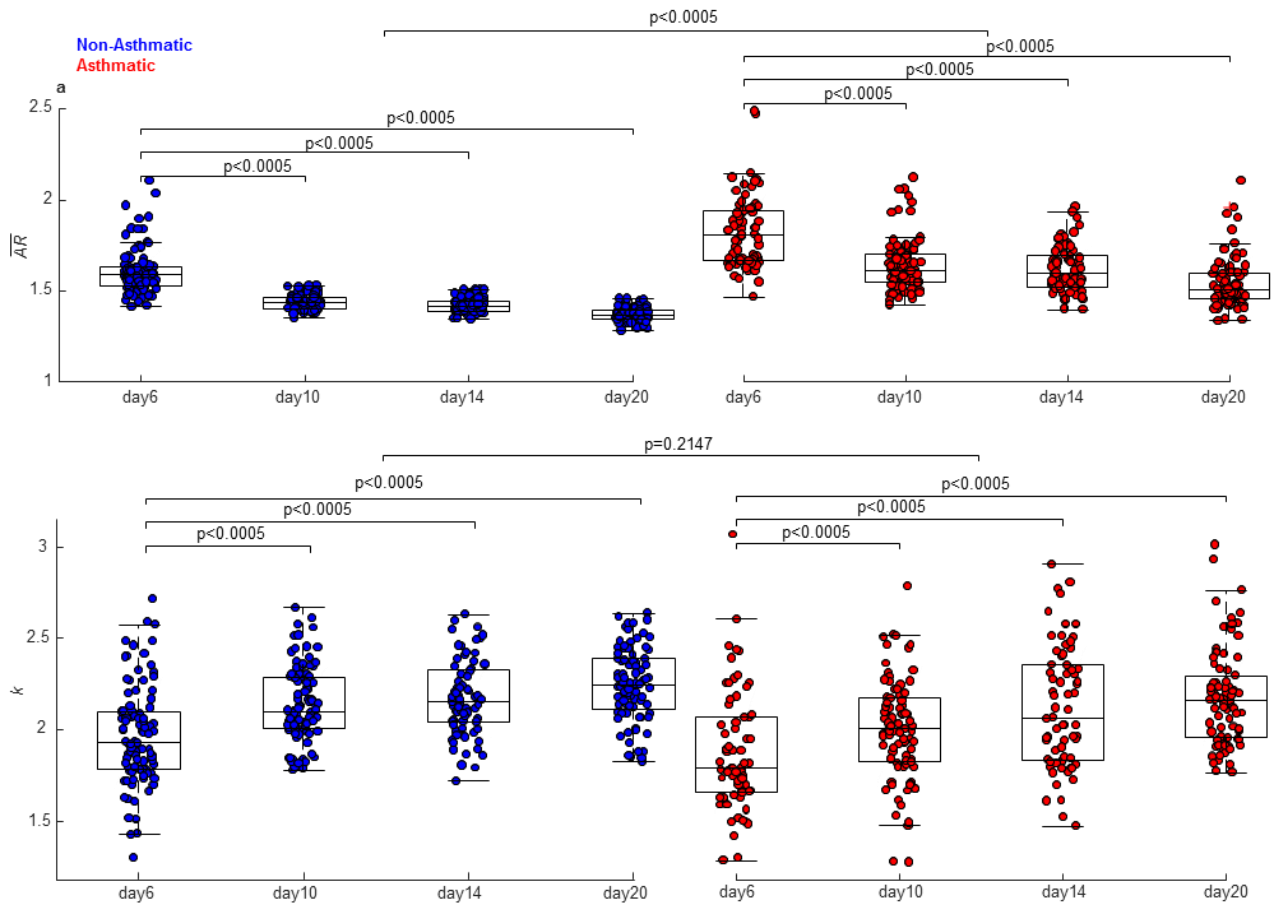

**Extended Data Figure 3| HBEC shape distributions were fully described by only two variables: $\overline{AR}$ and $k$.**
**a,** For both non-asthmatic and asthmatic HBECs, $\overline{AR}$ became progressively smaller with increasing days of maturation (p<0.0005), and across all days $\overline{AR}$ of the asthmatic HBECs was larger than for non-asthmatic ones (p<0.0005). (Each datum represents a different field of view).
**b**, PDFs of the rescaled AR (Fig. 3) followed the *k*-gamma distribution (Eq.1), analytically described by a single parameter, *k*. A common value of *k*=1.97 was observed across all days of maturation and across non-asthmatic and asthmatic HBECs (p=0.2147). (Each datum represents a fitted *k* value determined by maximum likelihood estimation, MLE, for each individual field of view).

**Extended Data Figure 4**

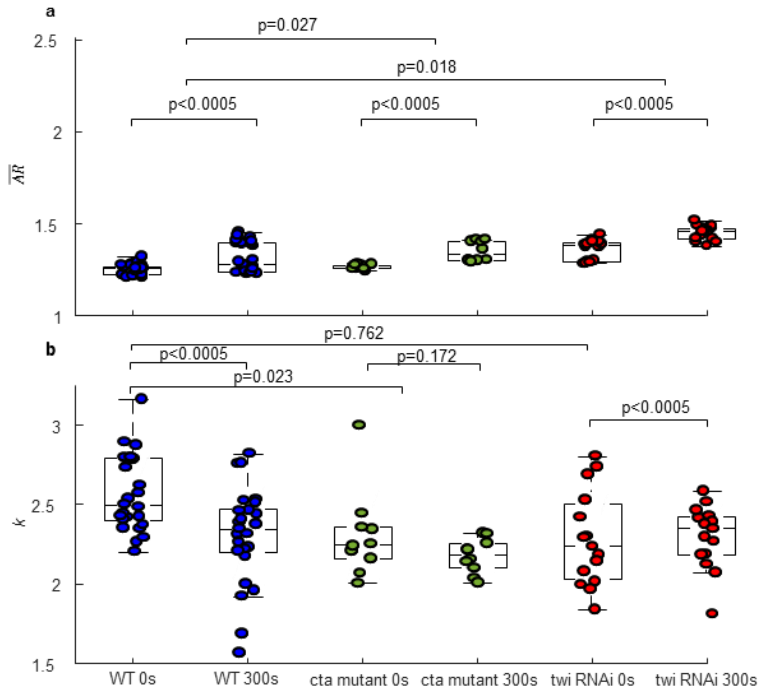

**Extended Data Figure 4| During ventral furrow formation in *Drosophila in vivo*, cells define a linear relationship between *AR* and its variability as well as a universal distribution of cell shape.**

**a,** For wild type (WT), *cta* mutant and *twist* (twi) RNAi embryos (n = 5, 2 and 3 respectively), $\overline{AR}$ increased with time (p<0.0005). Each datum represents a single frame from a sequence of five consecutive frames (~40s) in the vicinity of the specified time point (x-axis ticks).

**b**, PDFs of all rescaled AR data (Fig. 3) followed a *k*-gamma distribution (Eq.1), analytically described by a single parameter, *k*. (Each datum represents a fitted *k* value for a single frame). *k* slightly decreased with time in the cases of Wild Type and *twist* knockdown embryos (p<0.0005), yet a common value of *k*=2.52 (p=0.023) was observed across all.

**Extended Data Figure 5**

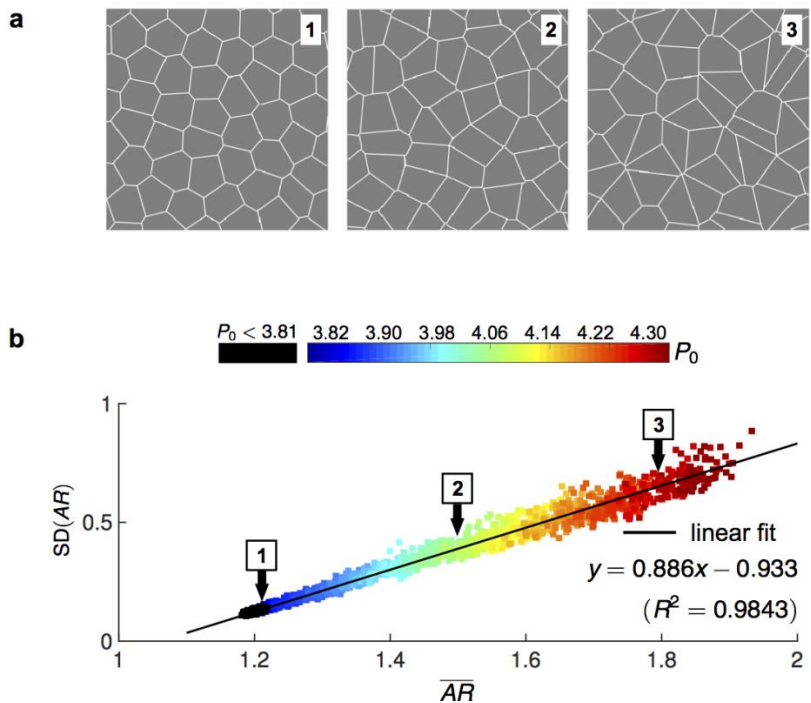

**Extended Data Figure 5|** Cell shapes and jamming in the self-propelled Voronoi (SPV) model. [27]

**a**, Typical simulation snapshots are shown, corresponding to $P_0 = 3.81$ (at the onset of jamming), $P_0 = 4.05$ and $P_0 = 4.25$ (both fluid-like).

**b**, Mean and SD of cell ARs from the SPV model are plotted for a large range of $P_0$ values. For $P_0 < 3.81$, the tissue is jammed where the mean aspect ratio (black line) remains nearly constant at 1.2. For $P_0 > 3.81$, the tissue unjamms and becomes more fluid-like. The standard deviation of AR is grows linearly with the mean and can be fitted to the linear relationship shown in the figure (black line). Labels 1-3 correspond to the AR values of the snapshots shown in **a**. The shear modulus of the tissue decreases as $P_0$ is increased and vanishes for $P_0 > 3.81$, indicating an ujamming rigidity transition[32,33].

# Supplemental Information

*1. Air Liquid Interface Culture*

The human conducting airways are lined by a pseudostratified layer of epithelial cells comprised of multiple cell types, including basal, goblet, and ciliated cells. These cell types are defined by location, function, and expression of cell type specific markers. The pseudostratified epithelium sits on and contacts with a basement membrane composing of collagens and lamins. Basal cells are progenitor cells of the airway epithelial cells and are identified by the positive staining of NGFR protein and by the expression of transcription factor, TRP63. Goblet cells and ciliated cells form the functional apparatus collectively referred to as the mucociliary escalator, in which pathogens are trapped in mucus secreted by goblet cells and cleared by the coordinated beating of cilia in the airways. Goblet cells are identified by the presence of mucin-containing granules positively stained for MUC5AC protein and by the expression of transcription factor, FoxA2. Ciliated cells are identified by the presence of cilia positively stained for beta-tubulin IV protein and the expression of transcription factor, FoxJ1. Culture in air-liquid interface of primary basal cells from human donors recapitulates the pseudostratified structure of the intact human airway epithelium, and is considered the gold standard in studying airway epithelial biology.

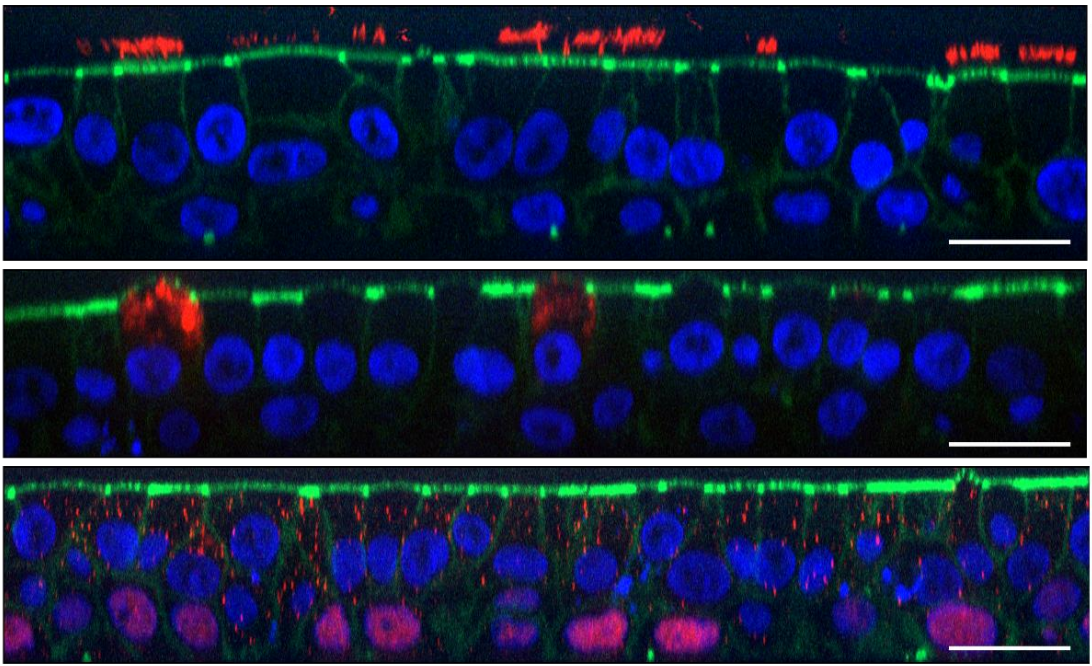

**Figure S1| Human bronchial epithelial cells form a pseudo-stratified layer.** Side-view images were reconstructed from z-stacks taken of HBEC layers fixed and stained on ALI day 20. In all three panels, green stains for F-actin, and blue stains for cell nuclei. In the top panel, the red color stains for β-tubulin which is a marker for ciliated cells. In the middle panel, red stains for Muc5AC which is a marker for goblet cells. In the bottom panel, red stains for p63 which marker for basal cells, which in some cases is found in the nuclei and thus leads to a pink color. Scale bars are 20 µm. The multiple positions of nuclei and cell features in these different kinds of cells lead to the appearance of a pseudostratified layer.

## *2. Theoretical Prediction and Simulations*

Here we simulate tissue configurations using a recent theoretical framework called the Self-Propelled Voronoi (SPV).[31] In the SPV model, the basic degrees of freedom are the set of cell center positions $\{\vec{r}_i\}$ and cell shapes are given by the a Voronoi of the point pattern $\{\vec{r}_i\}$. The complex biomechanics that govern intracell and intercell interactions can be coarse-grained[20,32,33,72] and expressed in terms of a mechanical energy functional for individual cell shapes. For a multicell tissue given by a collection of cell areas ( $A_i$ ) and cell perimeters ( $P_i$ ), the energy functional is given by

$$E = K_A(A - A_0)^2 + K_P(P - P_0)^2$$

The quadratic area term in the above equation accounts for a cell's resistance to volume changes via an area elastic modulus of $K_A$ and a homeostatically preferred cell area $A_0$[20].

Changes to a cell's perimeter are directly related to the deformation of the acto-myosin cortex concentrated near the cell membrane. The terms $K_P P_i^2$ corresponds to the elastic energy associated with deforming the cortex. The linear term in cell perimeter, $-K_0 P_0 P_i$, represents the effective line tension in the cortex and gives rise to a `preferred perimeter' $P_0$. The value of $P_0$ can be decreased by up-regulating the contractile tension in the cortex[20,32,33,72] and it can be increased by up-regulating cell-cell adhesion. In this work, we focus on the properties of the ground states of the SPV model and ignore the "Self-Propelled" aspect of the SPV model. We simulate a tissue containing N=400 cells under periodic boundary conditions with box size L = 20. The area modulus $K_A$ is set to 0 to simulate the large variation of cell areas observed in HBEC cells (Fig. 1). We also set $K_P = 1$. With these choices for the model parameters, the characteristic lengthscale in the system is related to the average area occupied by a cells $\langle A \rangle = L^2 / N = 1$, which is used as the unit length for all simulations. The reduced preferred cell perimeter $p_0 = P_0/\langle A \rangle$ is varied between 3.6 and 4.3. Initially, cell centers are randomly placed according to a Poisson point process and then evolved to energy minimum using the Broyden–Fletcher–Goldfarb–Shanno method.

It has been shown that by changing the value of the preferred cell perimeter $p_0$, tissues undergo a solid-fluid transition[32]. This transition occurs at a critical value of $p_0 = 3.81$. When $p_0 < 3.81$ the tissue behaves as a rigid solid with a finite shear modulus and there are finite energetic barriers for a cell to move or rearrange. For $p_0 > 3.81$ the tissue becomes a fluid, with a vanishing shear modulus (Extended Data Fig. 5).

The PDFs of the rescaled AR from the SPV model(Fig. 3g) can be well described by the k-gamma distribution (Eq.1). The value of k-parameter is obtained through least-squares fitting and displayed in Extended Data Fig. 5b. There is a weak but systematic dependence of the value of k on the parameter $P_0$. When jammed ($P_0 < 3.81$), k stays a constant at value of 2.4. However in the unjammed state, k decreases as $P_0$ is increased and appear to approach a lower plateau value of 1.8 in the limit of large $P_0$. The small value of k may be related to the fact that spatial correlations cell AR decay quickly as a function of distance between two cells. To check to see if this is indeed the case, we calculate two-point spatial correlations for the SPV model (not shown). We observe that regardless of distance to the jamming transition, cell aspect ratios are always short-ranged and never persist beyond 2-cell diameters.

## 3. Supplementary Table: Slope of Standard Deviation of AR against its mean

|  | Slope of SD(AR) vs AR | | | |
|---|---|---|---|---|
|  | slope | 95% lb* | 95% ub* | p-value |
| **Non Asthmatic HBEC** | | | | |
| Day 6 | 0.888 | 0.813 | 0.964 | <0.0005 |
| Day 10 | 0.913 | 0.788 | 1.038 | <0.0005 |
| Day 14 | 0.921 | 0.784 | 1.058 | <0.0005 |
| Day 20 | 0.799 | 0.666 | 0.931 | <0.0005 |
| **Asthmatic HBEC** | | | | |
| Day 6 | 0.717 | 0.660 | 0.773 | <0.0005 |
| Day 10 | 0.715 | 0.645 | 0.786 | <0.0005 |
| Day 14 | 0.710 | 0.621 | 0.799 | <0.0005 |
| Day 20 | 0.592 | 0.517 | 0.668 | <0.0005 |
| **MDCK** | | | | |
| All time | 0.870 | 0.860 | 0.880 | <0.0005 |
| *Drosophila* | | | | |
| Wild Type 0s | 0.9276 | 0.8726 | 0.9825 | <0.0005 |
| *cta* mutant 0s | 0.9977 | 0.8264 | 1.1690 | <0.0005 |
| *twist* RNAi 0s | 1.3147 | 1.1752 | 1.4542 | <0.0005 |
| *Drosophila:* **Change in slope per 100 seconds** | | | | |
| Wild type | 0.9280 | 0.9104 | 0.9448 | 0.960 |
| *cta* mutant | 1.0028 | 0.9807 | 1.0147 | 0.553 |
| *twist* RNAi | 1.3203 | 1.3000 | 1.3294 | 0.461 |

**Supplementary Table 1**| Slope of standard deviation of $AR$ vs its mean. In the first four groups of rows, p-value is the probability that the slope is zero; in the fifth group, p-value is the probability that the slope changes with time. Slope: the slope of SD($AR$) vs $\overline{AR}$; lb*: lower confidence bound; ub*: upper confidence bound.

## *4. Algorithm for identification of cells boundaries.*

To determine cells boundaries we used a semi-automatic segmentation pipeline. Intermediate results are depicted in Fig. S2. These included: **(a)** correction for non-uniform illumination by subtracting from the image its background, identified using *imopen* function in Matlab; boundaries were then enhanced according to the ratio between eigenvalues of local Hessian matrices [52,61,76,76,73,73]; **(b)** application of an intensity threshold limit of about 95% followed by a manual correction for disconnected boundaries; boundaries were then thinned and spur and isolated pixels were removed using *bwmorph* function in Matlab; the skeletonized image was then blurred using a Gaussian filter followed by a 99% intensity threshold; *bwboundaries* function in Matlab was then used to produce the final segmented image. **(c)** We defined a cell object boundary by the set of interior pixels which shares an edge with at least one exterior pixel, as also illustrated in Fig. S3.

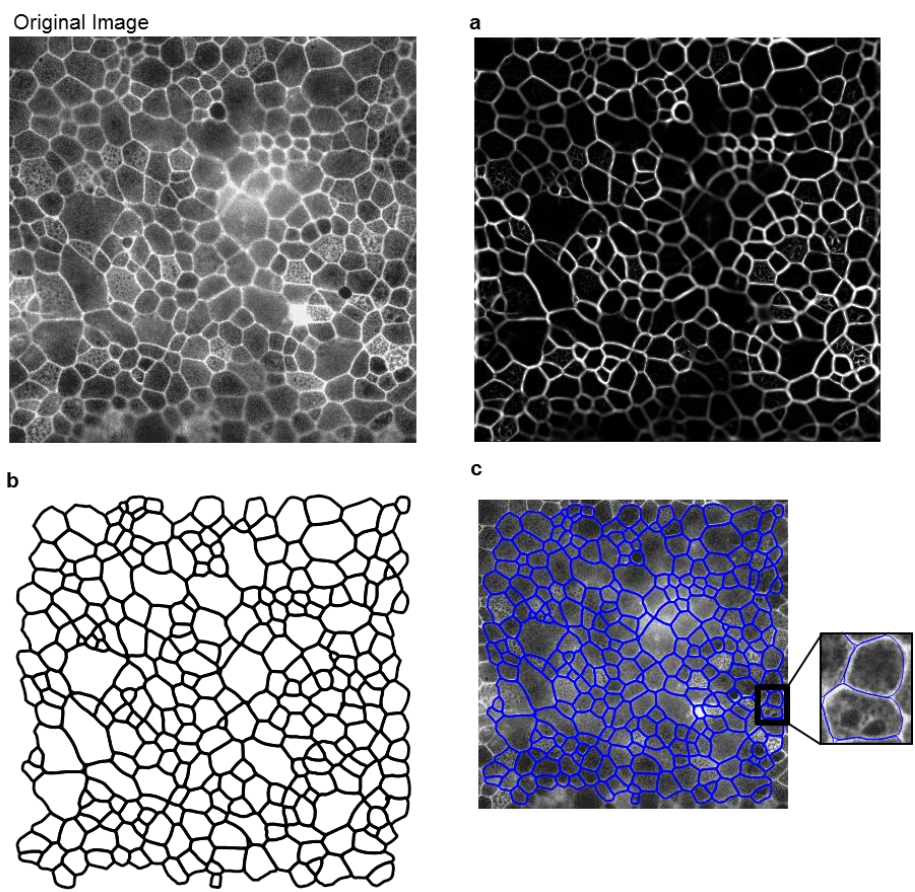

**Figure S2|** Identification of cells boundaries from an Epifluorescence image. **a**, Cell boundaries enhancement. **b**, Segmented image. **c**, Cell boundaries overlaid on the original image.

*Algorithm for shape measurement:* To measure cell areas and perimeters we adapted an approach taken by available commercial software. [53] For each identified cell $i$ in the final segmented image, the algorithm constructed a polygon whose vertices lie along cell boundary pixels (Fig. S3a). The polygon's $j^{th}$ side length is $1 \leq L_j \leq R_i$, with $R_i \equiv \max(0.1\sqrt{N_i}, 1.42)$, and $N_i$ is the total number of pixels composing the cell object. We defined the area and perimeter of cell $i$ as those of the polygon. The same polygon was used to evaluate the aspect ratio (*AR*) of cell $i$ as $\frac{a}{b} \geq 1$ (Fig. S3b), which is the ratio between the major and the minor axes of an equivalent ellipse, with equal eigenvalues of the second area moments as those of the polygon.
27

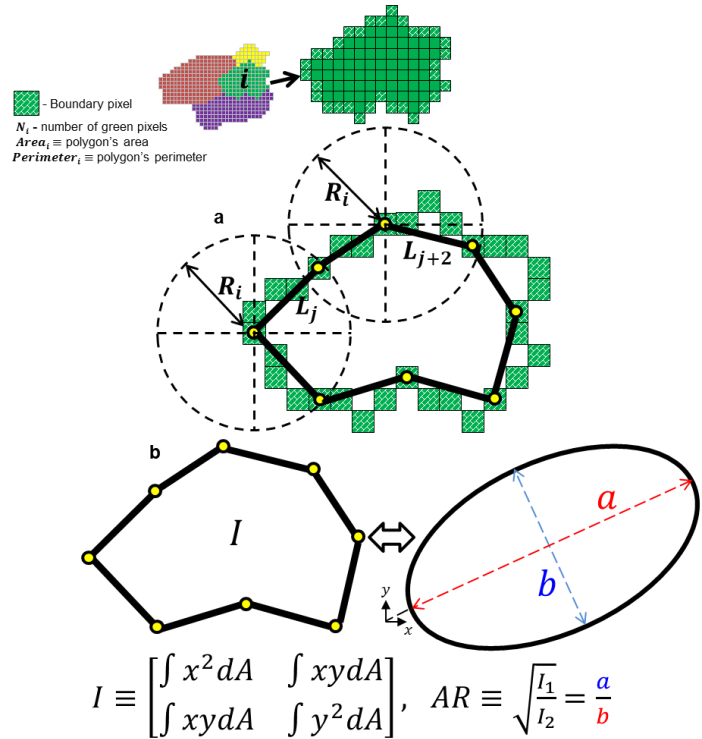

**Figure S3|** Shape and area measurements of cell *i*. Note that this is for illustration only, and a typical cell object is composed from thousands of pixels. **a**, Measurements of area$_i$ and perimeter$_i$ from a representative polygon with its vertices located along identified boundary pixels. **b**, An equivalent ellipse with equal eigenvalues ($I_1$ and $I_2$) of the second area moment $I$ as of the polygon's. The cell's aspect ratio (*AR*) is defined as $\frac{a}{b}$.

*Validation of algorithm performance:* We measured *AR* and *q* (perimeter/$\sqrt{\text{area}}$), both manually using imageJ software (National Institutes of Health) and automatically using the algorithm for shape measurement. Cell boundaries in blurry regions (Fig. S4b,c) were not properly identified. We attribute this to local height variations in the layer causing these regions to be out of focus. The accuracy of the algorithm was evaluated by comparing the histograms between the automatic and the manual measurements which showed closely similar shapes and an order of 1% difference in mean, median and standard deviation of both *AR* and *q* (Fig. S4d,e).

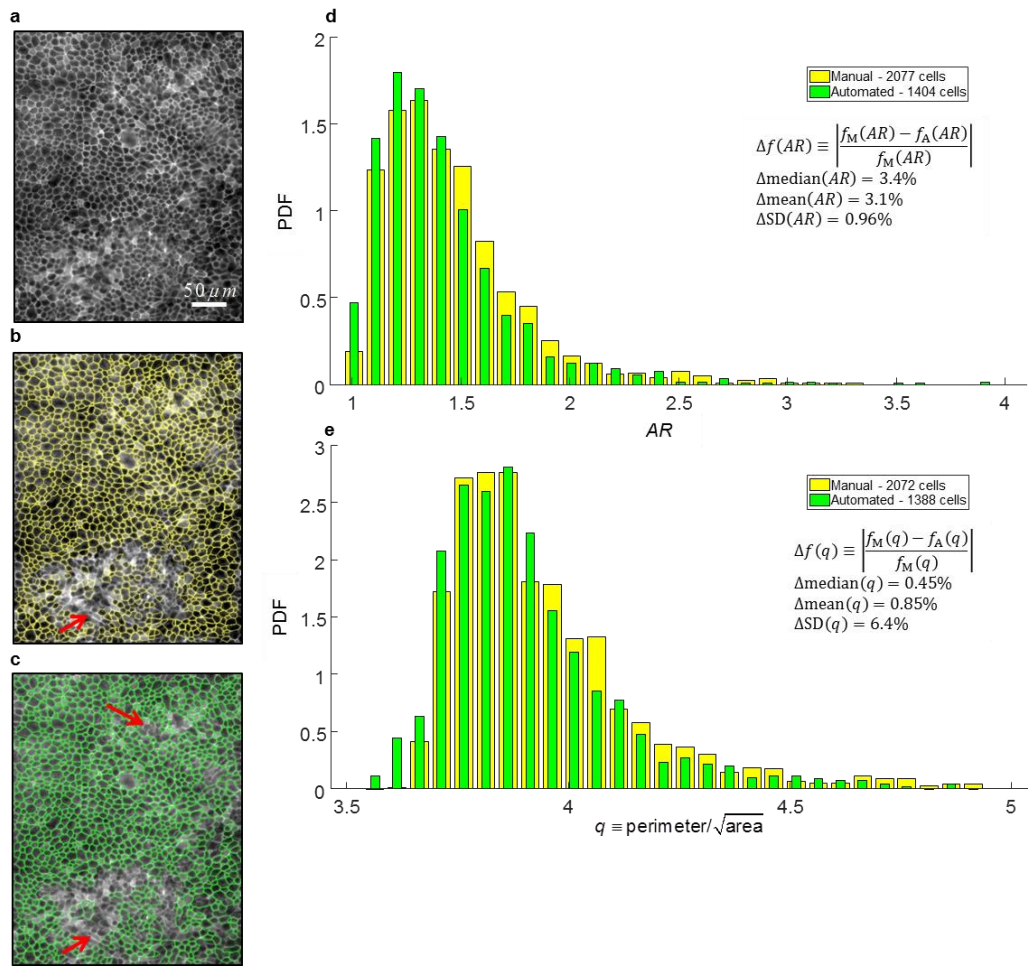

**Figure S4|** Automatic vs. manual detection of boundaries.

**a**, A typical Epifluorescence image of fixed primary HBECs labeled for cortical actin.
**b**, Manual tracing of boundaries in ImageJ overlaid on the original image.
**c**, Automatic detection of boundaries using our segmentation pipeline overlaid on the original image. Red arrows in b&c point to regions in which cells boundaries were not properly traced.
**d**, **e**, Comparing automatic vs. manual shape measurements. The histograms for *AR* (with bin size of 0.1) and for *q* (with bin size of 0.05) considers cell objects which were filtered such that $10\mu m^2 \leq$ Cell Area $\leq 2000\mu m^2$, $AR \leq 4$ and $q \leq 5$. The upper inset shows the resulted number of cells after filtration in each set. The lower inset shows the relative difference $\Delta f$ between different statistical moments $f_{M \setminus A}$ of the Automatic and the Manual measurements sets.
29

## 5. Discussion: Open questions and unresolved issues

The epithelial layers studied here are diverse, complex, and biologically relevant, but they are not to be confused with cell collectives that fully pack three-dimensional space. In that connection, the extracellular matrix (ECM) would be expected to provide local adhesive footholds for development of cell traction forces. But at the same time, ECM fibers would impose steric impediments to cell migration, screen force transmission from cell-to-cell, and induce cell channeling, all of which would be expected to impact cell jamming dynamics.[38,54,55] Although they might migrate on a basement membrane, the native tissues of central interest in this report –the human bronchial epithelial layer and the *Drosophila* embryonic epithelial layer– are otherwise constitutively devoid of ECM, as are many epithelial tissue compartments. Importantly, epithelia tend to be disordered and isotropic within in the cell layer plane, whereas, for example, the vascular endothelium subjected to shear flow becomes highly ordered and anisotropic [56], in which case geometric relationships as reported here would likely take a form that is quite different. In epithelia, unjamming and associated changes in cell geometries may represent a physical requirement for cell migration and cell rearrangements; to rearrange among themselves the cells within the layer must change shape and their distribution of shapes. But the relationship of unjamming to migratory events, the energetic efficiency of cell migration, and cell metabolism remain unclear. In HBECs, for example, proximity to the jammed state and the character of swirling motions have been shown to be tightly connected, and the tendency toward unjamming in asthma is suggestive of an injured or dysmature epithelial phenotype.[15] In cases such as the *cta* mutant and *twist* knockdown in the *Drosophila* embryo (Fig. 2), by contrast, cell geometries are clearly indicative of unjamming but collective migration is attenuated or unapparent, seemingly because formation of the ventral furrow has been impeded and associated tugging forces near the fold have diminished. In that connection, neighbor-swapping in a fully jammed collective is impossible, but the solid-like cellular collective might nevertheless be able to migrate *en bloc*, much as in the rotation of the *Drosophila* egg chamber driven by follicular epithelial cells.[57] Importantly, in the contexts of development, cancer, and the metastatic cascade, cells are thought to attain motility solely through the agency of an epithelial-to-mesenchymal transition (EMT)[58,59], but it remains unclear if unjamming might or might not fit within the EMT continuum. [58,60-63] The manner in which progenitor cell content, cell proliferation, cell extrusion from the layer, contact inhibition, and apoptosis might impact –and be impacted by– jamming dynamics are unknown.[34,64-68] Finally, it has yet to be established if the systematic geometrical features reported here might find practical application in examination of tissue specimens in clinical pathology.

## 6. Supplementary Videos

**Video 1 - https://www.youtube.com/watch?v=KB5ZWRVmGbM**

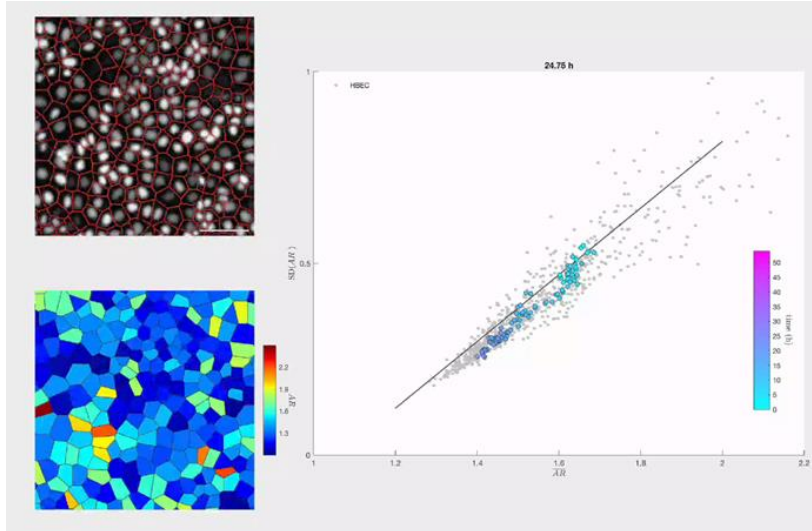

**Video 1| Proliferating Madin Darby Canine Kidney (MDCK) cells follow the same relationship as HBEC cells (Figure 1) with shape and shape variability mutually constrained.** $\overline{AR}$ decreased with time, but the relationship between $\overline{AR}$ and SD(*AR*) remained constrained. Top left panel: GFP nuclei with tessellated cells overlaid in red. Cells are only considered if resultant tiles have an area <1500 µm$^2$. Bottom left panel: corresponding color maps of *AR*. 10 to 64 hours after seeding are presented. Right panel: Each successive green datum represents $\overline{AR}$ vs SD(*AR*) with increasing time. Time as indicated, time 0 marks the beginning of the observation window. Gray: HBEC data. The black line is a theoretical prediction.

**Video 2 - https://www.youtube.com/watch?v=W6Rvi46GjWc**

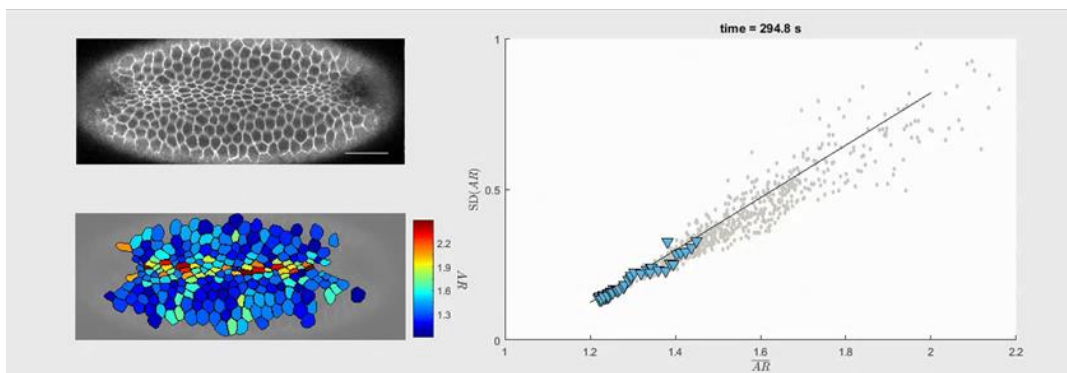

**Video 2| During ventral furrow formation in this WT *Drosophila* embryo, cell aspect ratio follows the same relationship as HBEC cells with shape and shape variability mutually constrained. This video accompanies data presented in Figure 2 (a-c, left column).** The values of $\overline{AR}$ vs SD(*AR*) follow the same relationship as HBEC, but in the opposite direction with time. Top left panel: z-projection of 3 consecutive slices imaging the ventral side of the embryo as cells constrict and attempt to form the ventral furrow (Fig. 2a,b left column). Data is presented until well before the furrow forms, when all cells are roughly in the same z-plane. Bottom left panel: corresponding color maps of *AR*. Right panel: Each successive blue datum represents $\overline{AR}$ vs SD(*AR*) with increasing time. Time as indicated, time 0 marks the beginning of the observation window. Gray: HBEC data. The black line is a theoretical prediction.

**Video 3 - https://www.youtube.com/watch?v=JmVRo8gKuFQ**

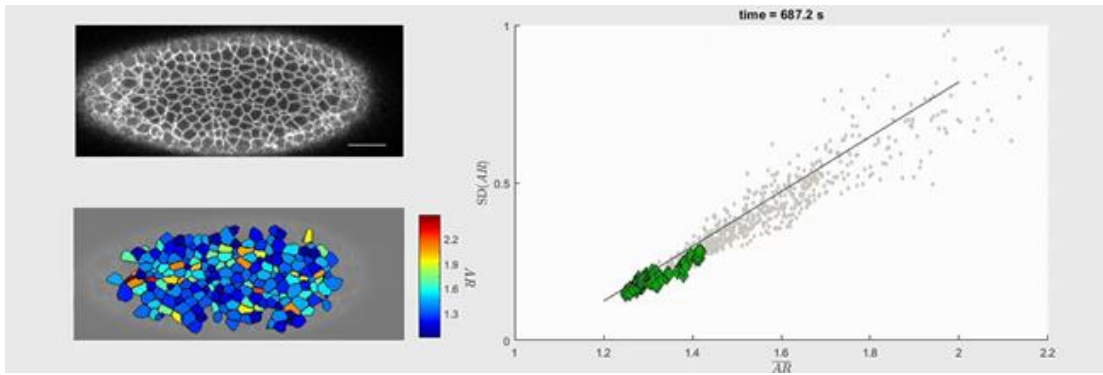

**Video 3| While cells in this *cta* mutant *Drosophila* embryo attempt to form the ventral furrow, cell aspect ratio follows the same relationship as HBEC cells with shape and shape variability mutually constrained. This video accompanies data presented in Figure 2 (a-c, center column).** The values of $\overline{AR}$ vs SD($AR$) follow the same relationship as HBEC, but in the opposite direction with time. Additionally, these data follow the same relationship as Wild Type embryos, even though furrow formation is defective. Top left panel: z-projection of 3 consecutive slices imaging the ventral side of the embryo as cells constrict (Fig. 2a,b center column). Data is presented when all cells are roughly in the same z-plane. Bottom left panel: corresponding color maps of *AR*. Right panel: Each successive green datum represents $\overline{AR}$ vs SD($AR$) with increasing time. Gray: HBEC data. Time as indicated, time 0 marks the beginning of the observation window. The black line is a theoretical prediction.

**Video 4 - https://www.youtube.com/watch?v=EAKPWUgGXL8**

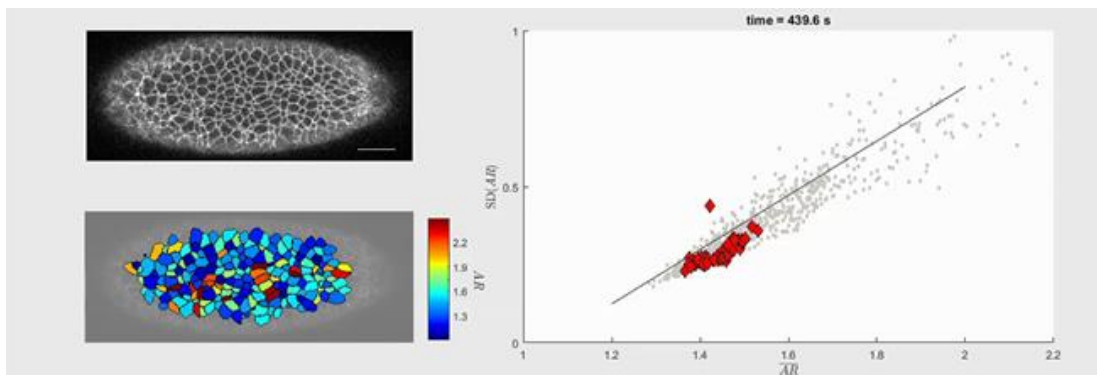

**Video 4| While cells in this *twist* RNAi *Drosophila* embryo attempt to form the ventral furrow, cell aspect ratio follows the same relationship as HBEC cells with shape and shape variability mutually constrained. This video accompanies data presented in Figure 2 (a-c, right column).** The values of $\overline{AR}$ vs SD($AR$) follow the same relationship as HBEC, but in the opposite direction with time. Additionally, these data follow the same relationship as Wild Type embryos, even though furrow formation is hindered. Top left panel: z-projection of 3 consecutive slices imaging the ventral side of the embryo as cells constrict (Fig. 2a,b center column). Data is presented when all cells are roughly in the same z-plane. Bottom left panel: corresponding color maps of *AR*. Right panel: Each successive red datum represents $\overline{AR}$ vs SD($AR$) with increasing time. Time as indicated, time 0 marks the beginning of the observation window. Gray: HBEC data. The black line is a theoretical prediction.

## 7. Supplementary References